# Kirchhoff's Current Law with Displacement Current

**Robert Eisenberg,** *Life Member IEEE*

*Department of Physiology and Biophysics, Rush University Medical Center, Chicago, IL; Department of Applied Mathematics, Illinois Institute of Technology, Chicago, IL; Department of Biomedical Engineering, University of Illinois Chicago, Chicago, IL USA*

**Xavier Oriols,** *Member IEEE*

*Departament d'Enginyeria Electrònica, Universitat Autònoma de Barcelona, 08193 Barcelona, Spain*

**David K. Ferry,** *Life Fellow IEEE*

*School of Electrical, Computer, and Energy Engineering, Arizona State University, Tempe, AZ 85287USA*

## I. INTRODUCTION

Ampere, Weber, Kirchhoff, Maxwell—all are giants whose work helped formulate our understanding of electrical circuit theory and electromagnetic waves. Certainly, three of these are recognized for their contributions to the understanding of electromagnetic phenomena. Kirchhoff, on the other hand, is recognized for his current and voltage laws in circuits, as well as his later work on black-body radiation from heated objects. Kirchhoff's current law is used widely to help design the circuits of our technology that respond in nanoseconds [1,2,3,4,5]. Kirchhoff's law has been used to design much slower circuits for nearly a century [6,7,8,9]. But, a casual search on the internet tells us that most people consider that Kirchhoff's Laws extend only to direct current (d.c.) circuits and do not fully include, for example, the displacement current extension necessary to treat high frequency circuits [10].

Kirchhoff's current law is about the currents that arise from the flow of charges in circuits—the flux of electrons—and the d.c. version does not deal with the rate of change of the total charge in the circuit. But the rates of change of charge and electric field are not small in circuits that respond in nanoseconds, and even in femtoseconds in pulse laser excitation of semiconductors [11] at one end of the physical scale and ion motion in biological channels at microscopic level [12]. The mechanisms and properties of current flow vary significantly between fractions of a femtosecond and seconds in wires and other systems [2,4,5]. If Kirchhoff's current law is limited to a form without time dependence, scientists are likely to have concerns about using it to construct physical understanding in systems where potentials change rapidly. Such concern is likely to increase further when they realize that Kirchhoff's d.c. current law is incompatible with the conservation law implied by Maxwell's equations, when explicit macroscopic time-dependent phenomena are involved.

Indeed, electrical phenomenon—slow (sec) and fast (nsec), even optical (fsec)—are described by Maxwell's equations [2,13] and do depend on the rate of change of the electric and magnetic fields. And, Maxwell is largely credited with the extension of Ampere's Law to these equations with his work of 1861, and the introduction of time-dependent displacement current [14]. But, is this view correct? Certainly, Maxwell presented a thorough discussion and rationale for adding displacement current, but the question to be asked is whether or not the concept was original to him? That is, is there evidence that this addition was known prior to 1861? In this paper, we point out that there is, in fact, evidence that Kirchhoff himself published a version that is thought to include displacement current additions to his own d.c. current laws, and this was done *several years earlier than Maxwell* [15,16].

In the following sections, the historical buildup to these papers of Kirchhoff and Maxwell will be discussed, as well as the importance of the displacement current in modern physical systems, from microwave circuits to biological structures. At the end, a discussion the importance of the work will be given.

## II. THE APPEARANCE OF DISPLACEMENT CURRENT

Beginning about 1820, and for the next five decades, the properties of magnetism and electromagnetics were formulated and entered the scientific domain. The names associated with this progress are among the most recognized in the electronics world: Ampère, Faraday, Weber, Kirchhoff, and Maxwell.

**A. Early Work**

Early in 1820, Ampère became aware of the work of Ørsted, who had observed that a current-carrying wire would deflect a compass needle [17]. Ampère carried the experiments further and established that two current-carrying wires would produce a force between them [18,19]. At the same time, he deduced that one wire produced an azimuthal magnetic field, and it was the interaction of this field with the second wire that produced the force between the two wires [20]. Of course, this result did not arise from a single measurement. Ampère carried out a series of measurements, varying the angle between the two wires (wires perpendicular to one another had no force between them), using wire rings and spirals, among other designs. One of these was the solenoidal wire, which he found provided a magnetic field similar to that of a bar magnet. Ampère would summarize his work in currents and force in a memoir published a few years later [21]. Maxwell would later refer to this work as [22]: "The experimental investigation by which Ampère established the laws of the mechanical action between electric currents is one of the most brilliant achievements in science." Certainly, Ampère's work laid the foundation for the advances and investigations that followed for the next several decades, that culminated in Maxwell's equations.

Only a few years after Ampère's critical work, Faraday extended the effects by noting that moving one of the wires, or varying the current in the other wired, increased the heating of the wire apparently through an increase in the current in the moving wire [23]. He concluded that the time dependent interaction with the magnetic field induced a potential (or "tension") in this wire that led to the added current, and this was the beginning of magnetic induction. From this, it became clear that time variation was now an important aspect of electric circuits, but only so far as the magnetic field was involved. In further studies, he clearly showed that one could think of magnetic field lines, and that these were closed quantities, having no beginning or end. The flux quantity of such a line would be the same within a solid as outside the solid [24]. Faraday also noted that the moving wire gained the maximum inductive effect when moving perfectly transverse to the magnetic field. Again, he reinforced the importance of time on the dynamic properties of the fields [25], coming over the various periods to regard both the static and dynamic effects with the phrase of electro-tonic state.

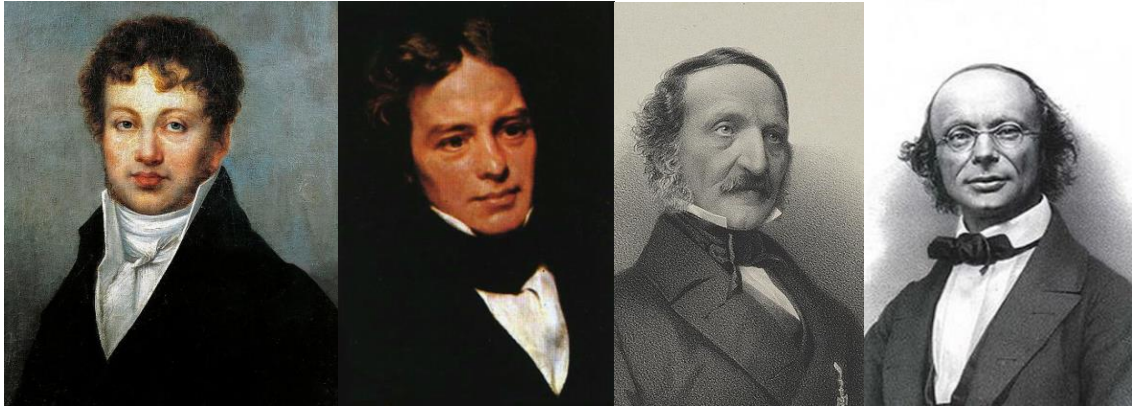

Fig. 1. Pioneers in the development of electricity and magnetism. From left: André-Marie Ampère, Michael Faraday, Franz Ernst Neumann, and Wilhelm Weber.

Wilhelm Weber would continue the study of electromagnetic voltage induction, as he would call it [26]. Weber pointed out that, other than the work of Faraday, not much had been done since Ampère's original work in the field of electrodynamics. He proposed more elaborate methods of making the experiments, and would ultimately put forward the conditions for absolute determination of the currents, voltages and magnetic fields. These procedures would ultimately provide the standardization of such measurements in many countries. But, he continued to call the charge (as we know it today) as the electrical mass. He also showed that in the case of static currents and wires, the electrodynamic effects would vanish, so that the latter were intimately involved with time variation, either of the position of one of the wires or by variation of the current. His view was that the whole theory of electricity contained both electro-statics and electrodynamics. Nevertheless, his electrodynamics did not yet include the results of time varying voltages in the absence of magnetic effects. This addition was still to come.

By 1855, James Clerk Maxwell had finished his undergraduate work and had become a Fellow at Trinity College, Cambridge. Already as an undergraduate, he had begun his scientific career, delving into many fields of mechanics, but turning to electrodynamics only at this time. This was a paper on Faraday's lines of (magnetic) force [27]. As a novel approach, he introduced a fluid analogy to the lines of force, and showed how much of the physical properties could be explained in this manner. Although he treated several examples, Maxwell considered the fluid to be incompressible, which precluded the possibility of considering charge accumulation in the system. Nevertheless, he did consider the properties and theories of dielectrics as well as magnetic materials, noting that Thomson had also written about magnetic induction [28]. And, he gave some discussion of multiple current carrying wires, although he did not appear to be aware of Kirchhoff's theorems as yet.

## B. Kirchhoff on Currents

Gustave Kirchhoff was still a student in Königsberg when he first developed his ideas on the summation of currents and voltages. He apparently gave a talk on his ideas at a monthly meeting of the academy of science in late 1844, and the gist of the talk was published in a short report by the editor of *Annalen der Physik* [29]. Some of this effort was inspired by Neumann, who was studying Faraday and Weber's work on induced currents [30]. Kirchhoff would follow this with two papers of his own on the topic [31,32]. He continued to work on the problem through the years, although at this time he was still considering only the d.c. situation [33]. In 1847, he moved to Berlin, but continued to extend the work to formulas for linear, as well as non-linear, ladder networks [34]. He then turned to induced currents, following the forementioned work [35,36]. Following this work, Kirchhoff moved to Breslau and then to Heidelberg, and for some time focused his studies on optical spectroscopy, working with his colleague Bunsen. But, he would return to electricity.

It is important to note that researchers at the time considered that the current in the wire resulted from positive electricity and negative electricity (we would call these charges today). It is not clear when this distinction came into use, but it is clear that Kirchhoff discussed the currents in this manner. At the time, it was considered that these two components would be equal, both contributed to the current, and this led to extra factors of 2. When Kirchhoff returned to electricity in Heidelberg, he asked the question about what would happen if these two components were not equal. These studies led him to two important results: (1) that the voltage in a circuit was related to the difference in the two electricities (and reached an early version of what today is known as the scalar Liénard-Wiechert potential [37,38], found by these two five decades later), and (2) an equation that relates

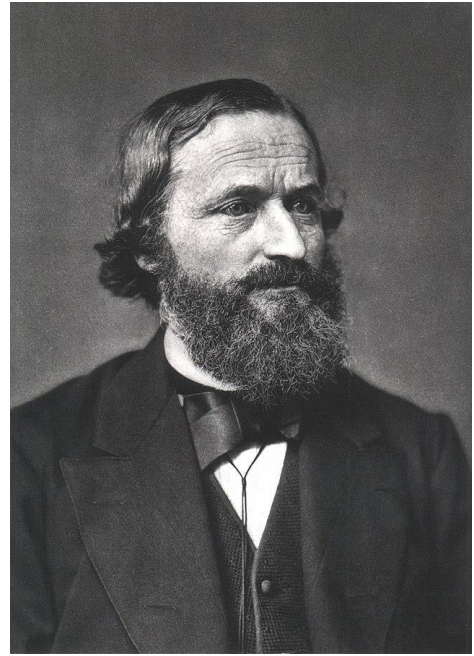

Fig. 2 Gustav Robert Kirchhoff.

the current density to the difference in the two electricities, called the *free* electricity [15]

$$2\frac{\partial i}{\partial s} = -\frac{\partial e}{\partial t}, \tag{1}$$

where the fact of 2 arises from the earlier discussion on currents, *s* is the distance along the wire and *e* is the difference between the positive electricity and the negative electricity. Graneau and Assis point out that this leads immediately to the continuity equation [16]

$$\nabla \cdot \boldsymbol{J} = -\frac{\partial \rho}{\partial t}, \tag{2}$$

in modern notation and omitting the factor of 2. In his second paper of the year, Kirchhoff rewrites (1) in the form [39]

$$2\frac{\partial i}{\partial x} = -\frac{\partial E}{\partial t}, \tag{3}$$

where *E* is described as the quantity of free electricity per unit length. If this is interpreted as related to the modern flux density *D*, then a version of displacement current has arrived.

The first version, given in (1), also appeared in English, as it was almost immediately reprinted in the British journal mentioned in [15]. As this was one of the primary journals at the time, it is unlikely that Maxwell would not have seen the paper. However, the second paper does not seem to have been translated, but Maxwell published earlier papers in both German and Italian, so it may be supposed that he likely read this paper as well.

The importance of (1) and (2) lies in the modification that must be made to the d.c. form of Kirchhoff's current law. Consider the standard approach of enclosing a circuit node within a spherical shell/volume, and then applying (2) within that volume. This is done by integrating over the volume of the sphere, and using the divergence theorem to change the first term into a surface integral, as

$$\int \boldsymbol{J} \cdot \boldsymbol{n} dS = -\frac{d}{dt}\int \rho d\Omega = -\frac{dQ}{dt}. \quad (4)$$

Here, the surface normal points outward and $\Omega$ is the volume of the sphere, with *Q* being the total charge within the sphere. If the surface integration merely sums over the currents in the wires connected to the specified node, then this last equation becomes

$$\sum_i I_i = -\frac{dQ}{dt}. \quad (5)$$

Hence, this becomes the a.c. form of Kirchhoff's current law—the sum of the currents *leaving* a node is accompanied by a reduction in the free charge at that node.

It is hard to come to the understanding of how Kirchhoff came to (1) and the role of free electricity. He did not leave a series of papers between 1849 and 1857, so ascertaining the thought process is difficult. He was working on other topics during this period, but perhaps the idea of compressibility of the electricity occurred to him. If, for example, the negative electricity is composed of negatively charged particles, as is now known, then this would lead to a compressible "gas" of particles. Recognizing this would lead the realization that the positive electricity did not need to actually balance the negative electricity locally (although it did need to do so globally over the device or system). This would then lead to the concept of free electricity and lead to (1) and (2).

**C. Maxwell Comes to Displacement**

By 1860, Maxwell had been and gone from Aberdeen and was ensconced at Kings College, London. The next several years are generally considered to be some of his most productive time. At the beginning, he returned to consideration of the lines of force in electromagnetics [14]. In this work, he brought forward the idea of a network of small vortices that were important in magnetism and he thought they would be important in electrostatics. But, he also brought forward the idea of the continuity equation. Here, he considered *p*, *q*, and *r* to be the electrical currents in the x, y, and z directions, and *e* to be the net free electricity per unit volume (now our charge density ρ). Then, the continuity equation may be written as

$$\frac{\partial p}{\partial x} + \frac{\partial q}{\partial y} + \frac{\partial r}{\partial z} + \frac{\partial e}{\partial t} = 0. \quad (6)$$

While Maxwell does not reference Kirchhoff, he uses the same expression of free electricity and this last equation is the three-dimensional version given by Kirchhoff as (1) (neglecting the extraneous factor of 2). He also related the charge density to the electromotive forces, which led to lines of force that begin and end on the free charge. Thus, if *P*, *Q*, and *R* are the forces along the three axes, then the connecting law is given as [14]

$$e = \frac{1}{4\pi E^2}\left(\frac{\partial P}{\partial x} + \frac{\partial Q}{\partial y} + \frac{\partial R}{\partial x}\right), \quad (7)$$

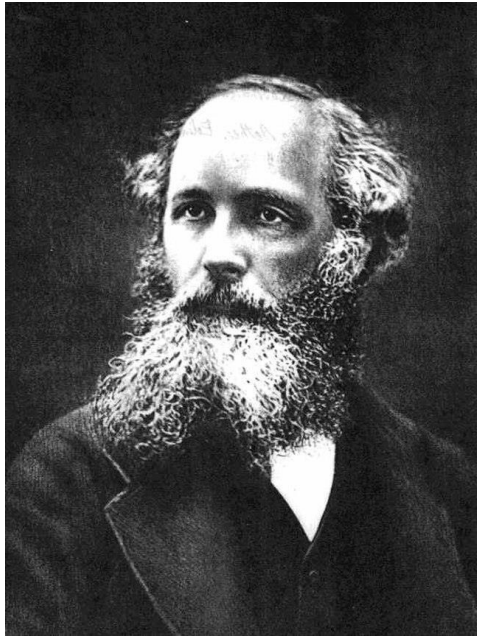

Fig. 3. James Clerk Maxwell.

where $E$ describes properties of the dielectric in which the forces exist.

Maxwell would go beyond this work with his later paper on the equations of electromagnetics [40]. Here, he discussed the fact that free electricity would pile up at the ends of a volume in which it occurred. This would provide a displacement between positive and negative electricity. But, he also noted that in dielectrics, individual molecules of the material would respond to applied electromotive forces in a manner which gave a polarization of these molecules in response to the electromotive forces. He proposed that in response to the electromotive force, the molecules would have positive and negative ends, and these would align in the force to create the polarization. He was clear that this displacement, in either case, did not produce currents, but they would produce currents as the polarization increased or decreased. Using his previous notation for currents, and $f$, $g$, and $h$ as the displacements in the three directions, the modified currents $(p', q', r')$ would be related to these displacements as

$$p' = p + \frac{\partial f}{\partial t}$$
$$q' = q + \frac{\partial g}{\partial t}. \quad (8)$$
$$r' = r + \frac{\partial h}{\partial t}$$

He extended the work to show that the rotation of the magnetic forces would be driven by these modified currents. These were basic tenets of his new theories on the electromagnetic fields. Nevertheless, it has to be said that the work of Kirchhoff presaged much of this development, although Maxwell certainly went well beyond Kirchhoff's earlier work.

### III. WHY DOES IT MATTER

Kirchhoff's current law is about the flow of charges in circuits, often the flux of electrons, and did not originally deal with the rate of change of the total charge in the circuit. The rate of change does not appear in a term in the usual formulation of Kirchhoff's (d.c.) current law. But the rates of change of charge and electric field are not small in systems that respond on the short-time scale. The mechanisms and properties of current flow vary significantly on the nanosecond (and shorter) scale. Yet, we have shown that Kirchhoff himself modified the equations to account for time varying "free electricity" in (1). The importance of this lies in Maxwell's extension of Ampère's law, in his (using modern notation,

$$\nabla \times \mathbf{H} = \mathbf{J} + \frac{\partial \mathbf{D}}{\partial t}. \quad (9)$$

Here, the right-hand side has been modified as from (8). $\mathbf{H}$ is the magnetic field intensity, and for a great many years was measured as so many lines per unit length (in the English system), in keeping with Faraday's lines of force [25]. Today, with the m.k.s system, it is measured as Amps/m. The quantity $\mathbf{D}$ is known as the electric flux density, measured as Coul./m$^2$. In coming to this flux density, Maxwell assumed that the free electricity would ultimately reside on the surface of the material, but would also include the polarization of molecules that made of the material. On the other hand, Kirchhoff noted that this need not be the case [15], but free electricity could be completely internal to the material, such as in *p-n* junctions, or even outside the material in cases such as time-dependent electron emission, or in optics where there is no particle current flow (it is well known that one needs the second term on the right-hand side of (8) in order to arrive at the wave equations for the scalar and vector potentials in high frequency electromagnetics [2]). Today, **D** is known to account for all of this with the relation

$$\mathbf{D} = \epsilon_0 \mathbf{E} + \mathbf{P}, \qquad (10)$$

where **E** is the electric field, **P** is the polarization in the medium, and $\epsilon_0$ is the permittivity of free space.

Conservation of current involves the displacement term because conservation is described by the divergence operator that can be applied to both sides of (9) to give

$$\nabla \cdot (\nabla \times \mathbf{H}) = 0 = \nabla \cdot \mathbf{J} + \frac{\partial (\nabla \cdot \mathbf{D})}{\partial t}, \quad (11)$$

which becomes (2) once one applied Maxwell's constitutive equation [2]

$$\nabla \cdot \mathbf{D} = \rho. \qquad (12)$$

*It is the total current, not just the charge flow current, that is conserved in general.*

There are other conditions needed to make Kirchhoff's current law 'exact', and these are often difficult to define broadly in mathematical form, because they arise from defining a connection between the polarization and the flux density in a wide range on quite different systems. Other sufficient requirements and conditions are not apparent in the diagrams of circuits that are analyzed with Kirchhoff's laws, yet the additional requirements may be important, especially when they depend on the properties of components and the location of the stray capacitances that link everything, including structures outside the circuit itself [5]. In a sense, all of this depends to a large extent on how the polarization is related to the flux density, and this depends upon the properties of the material as well as the physical layout of the system.

In the linear response world (that is, where **E** and **D** are small), one can write the polarization as

$$\mathbf{P} = \chi_e \epsilon_0 \mathbf{E}, \qquad (13)$$

where $\chi_e$ is the electric susceptibility. Using this equation in (10), one may define the relative permittivity as

$$\epsilon_r = 1 + \chi_e, \qquad (14)$$

and a linear relation between **D** and **E** results. But, this linear relation only holds in the linear response regime when the "molecules" respond linearly to the electric field [41]. This is an important point, as this response of the molecule is a separation of the positive (nucleus and core electrons) and the bonding electrons in the material. These are all particles of one form or another. Even in linear response, however, there are system whose complexity is such that a *single* relative permittivity cannot be defined due to complicated spatial or temporal behavior. Any values must be specified as to their particular environments, especially with respect to the frequency of interest. This is especially true when studies of photo-reflectance, ellipsometry, or other spectroscopy is being utilized [42,43,44,45].

In the linear response used above, **P** is described by a simple dielectric model, but it must be remembered that it can actually be strongly dependent on space and time. In conducting media, the susceptibility can even be complex. There are other cases where this approach will never hold, such as the permanent polarization in ferroelectric materials. Here, one cannot really define a relative permittivity. There are other cases where the relationship between **P** and **E** is nonlinear. Even in Si, the relationship between polarization and the electric field is sufficiently nonlinear [46] to be used for the important technology of nonlinear optics, either classical or quantum mechanical. The existence of this nonlinear relationship means that one must be very careful about the conditions for which a relative permittivity is used [47].

There is an additional constitutive relationship between the magnetic field intensity **H** and the magnetic flux density **B**, given (in free space) by

$$\mathbf{B} = \mu_0 \mathbf{H}, \qquad (15)$$

where $\mu_0$ is the permeability of free space. In real material, this linear response is brought into question, in particular in magnetic material where a magnetization (analog to polarization) appears and changes the response as

$$\mathbf{B} = \mu_0 (1 + \chi_m) \mathbf{H}, \qquad (16)$$

where $\chi_m$ is the magnetic susceptibility. Like with ferroelectrics, there are ferromagnetic material where there is a permanent magnetization. But, linearity is a difficult property to obtain. In both ferroelectrics and ferromagnetics, the polarization and magnetization, respectively, can be reversed by the driving field (**E** or **H**, respectively). This switching is accompanied by hysteresis, which is never linear in behavior! Indeed, this hysteretic behavior can even be "observed" in some materials which are definitely not ferroelectric [48]! This unusual hysteresis has also suggested a pseudo-photovoltaic response [49].

## IV. SOME IMPLICATIONS

The importance of the displacement current is well recognized today in both electromagnetics and circuits, particularly for the a.c. case. The obvious exhibit is the capacitor, where no d.c. current can flow through it, but displacement current through the insulator balances the wire current due to particles in the wire, as shown in Fig. 4. The dielectric in this case is free space (vacuum), but in real capacitors there is a dielectric material that is characterized by a relative dielectric "constant." Perhaps the most common, in terms of number of capacitors made, is the metal-oxide-semiconductor capacitor in DRAM. This is perhaps a pertinent example, as the dielectric polarization in the oxide is extremely nonlinear, with singularities at two frequencies corresponding to far infrared. These nonlinearities will crop up in fast switching of these capacitors. But, there are more situations where displacement current can be both more difficult and more meaningful.

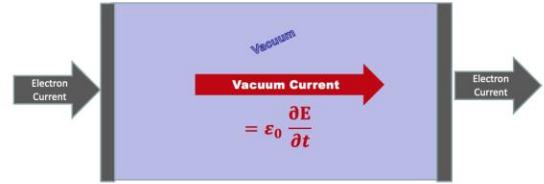

Fig. 4 A vacuum capacitor. In a normal capacitor, the permittivity would be corrected with the relative permittivity of the material used as dielectric in the vacuum region.

### A. Waves

To express this, the additional Maxwell equation

$$\nabla \times \mathbf{E} = -\frac{\partial \mathbf{B}}{\partial t} \qquad (17)$$

is needed. To proceed, the curl of (9) is taken, in the approximation of an insulating material, and then (17) is used in the right-hand side to give

$$\nabla \times (\nabla \times \mathbf{B}) = -\mu\epsilon \frac{\partial^2 \mathbf{B}}{\partial t^2}, \qquad (18)$$

where (13, 14, 16) have been used to introduce the permeability and the permittivity in the linear approximation. Expanding the double cross product, and using the additional Maxwell equation $\nabla \cdot \mathbf{B} = 0$, leads to

$$\nabla^2 \mathbf{B} - \mu\epsilon \frac{\partial^2 \mathbf{B}}{\partial t^2} = 0 \ . \qquad (19)$$

This is simply the normal wave equation that results for all wave fields in the linear response approximation. Note that it cannot be derived without the displacement current term—it is absolutely crucial to this result. But, this result hides some uncomfortable truths.

Suppose the order of equations is reversed, and (10) is used. Then, (19) becomes, for the electric field,

$$\nabla^2 \mathbf{E} - \mu\epsilon_0 \frac{\partial^2 \mathbf{E}}{\partial t^2} = \mu \frac{\partial^2 \mathbf{P}}{\partial t^2} \ . \qquad (20)$$

Of course, this reduces to the same wave equation as (19) when linear response is assumed. However, linear response is a very special case, and represents only a very small part of the world of electromagnetics. Certainly, the general case is that $\mathbf{P}$ is time varying, just as $\mathbf{E}$ is. If $\mathbf{P}$ is also either nonlinear in the field or is inhomogeneous, the result is not simple wave propagation, but can lead to very complicated nonlinear equations and/or distinctly different propagation properties in different crystalline directions (within a crystalline material) [50]. The entire field of nonlinear optics depends upon moving beyond linear response. The fact that there is so much effort (and publications) in microwave theory and techniques unfortunately masked the point that it is based upon a relatively simplistic approximation.

A particular example of the difficulties is the millimeter integrated circuit (MMIC). Transport of the millimeter waves on the MMIC is usually by strip lines (open waveguides which induce propagating guided by a top surface metal strip line and the underlying ground plane, using a non-absorbing substrate material), although coplanar waveguides are also used [51]. These waveguides have relatively low impedance (lower than free space), but must be matched to the very-high impedance reactive inputs and outputs of the transistors. This requires complicated matching networks to be included in the circuit. Moreover, the millimeter waves must be isolated from the power leads, and the d.c. power must be isolated from the waveguides. All of this requires design constraints which are not always compatible.

**B. Ion Channels**

An important application of the conservation of total current is in ion channels of biological membranes and nanotechnology. These ion channels are mostly narrow pores through proteins that allow otherwise impermeable ions to pass into cells. Ion channels control an enormous range of biological function in health and disease and are extensively studied. The narrow pores of biological channels are rarely wide enough to allow ions to pass by each other with high probability. The current flow through the pores has been viewed as a single file hopping phenomenon [52,53]. The ion channels of nerve, skeletal, and cardiac muscle responsible for nerve signaling and the coordination of contraction use total current to make the nerve signal, as is apparent from both experiments and theory [54]. Here, it is important to recognize that nerve signals are not d.c., but are pulsed a.c., and require displacement current for their efficient signal propagation.

The single file passage of the ions certainly is of great importance for the charge current carried by these ions through the channel, as shown in Fig. 5 [55]. But the total current through the pore of the channel includes another component, the displacement current produced by the polarization of matter and space. The sum of those components is conserved even though the individual components are not. The charge current varies dramatically with position. The displacement current varies dramatically with position. But their sum does not vary with position, at all, as indicated by (11).

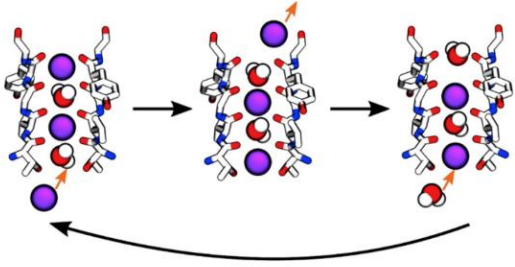

Fig. 5. Serial passage of potassium ions through an ion channel. The black just denotes repeating the sequence of conformations. Reprinted from A. Mirenenko et al. [55], under the creative commons usage license.

Maxwell's equations—and versions of Kirchhoff's law that are consistent with these—ensure that total current is conserved whenever these equations are used [56]. In a narrow single file channel, the displacement current takes over from the charge current (and vice versa) exactly so the total current is constant along the length of the narrow channel. The consequences of the interplay of charge current and displacement current is to simplify the system dramatically. The total current does not vary with spatial location in a narrow one dimensional system. However complicated are the hopping and single file behaviors, the total current is the same at all spatial locations in the channel because one component of the total current takes over from the other, to make it so, as Maxwell's equations require. The electric and magnetic fields change the movement of charges on the atomic scale to make this so.

The implications for atomic scale theory were clearly known in theories of one dimensional transport [56]. In other words, a theory of the total current does not need to have the spatial location as an independent variable. Of course, a theory of total current is not a complete theory of electrodynamics, let alone charge movement. The spatial variable is obviously needed for complete understanding. In many situations, however, a measurement of total current is enough to allow significant understanding and control of a system. Those situations include many of the circuits of our electronic technology. They also include many ion channels.

### C. High Frequency Quantum Devices

The great success of our information society is based on encoding the physical values of currents and voltages inside electron devices as digital (or analog) information. Typically, the simulation of such devices is done considering only the particle current, while ignoring the displacement current. But, this fails at high frequency as noted above. To understand when this low frequency assumption is acceptable, consider some values for the total current $\mathbf{J}_T$ mentioned in (9) and (11) as

$$\mathbf{J}_T = \mathbf{J} + \frac{\partial \mathbf{D}}{\partial t} \approx \sigma \mathbf{E} + i\varepsilon\omega\mathbf{E} \quad . \qquad (21)$$

On the right-hand side of (21) the particle current is proportional to the electric field through the conductivity $\sigma$ and the displacement current is evaluated assuming a sinusoidal temporal dependence of the displacement field $\mathbf{D} = \varepsilon\mathbf{E}$ corresponding to a frequency ω. The typical conductivity $\sigma$ in Silicon is less than $10^{-1}\ \Omega^{-1}m^{-1}$ and, using $\varepsilon \approx \varepsilon_0 = 8.85 \cdot 10^{-12}\ Fm^{-1}$, the displacement current in semiconductors devices can be safely ignored up to few hundred GHz. However, since the displacement current in (21) grows linearly with ω, the displacement current cannot be ignored for high enough frequencies (obviously, the displacement current cannot be ignored at any frequency in capacitors because the conductivity is zero, as discussed above).

How displacement current is modeled in semiconductor devices working at hundreds of GHz is the question? A straight-forward answer comes from the semi-classical simulation of electron devices. For example, the typical Monte Carlos solution of the Boltzmann equation provides the semi-classical trajectory $x(t)$ for each electron so that the total charge density $\rho$ can be defined. Then, the displacement current in (21) can be evaluated from the time-derivative of **D** obtained by using Gauss' Law in (12). But, when quantum phenomena become relevant, the mandatory inclusion of the displacement current in quantum transport simulators becomes a more complicated issue, either from a computational or fundamental point of view [57,58].

According to the orthodox quantum theory, any measured property of a system coincides with the eigenvalue of an operator linked to such property, and the state of the system "collapses" into the eigenstate of such eigenvalue. For modeling DC currents, the "collapse" is ignored assuming that time-averaged current is equivalent to an average over identical devices whose current is measured just once. However, the previous ergodic argument is no longer valid in far-from-equilibrium semiconductor devices, especially at high frequencies. In principle, then, one would have to face the perplexing effects of the "collapse" postulated by the orthodox machinery. However, in practice, this orthodox theory is avoided by more causal versions of quantum mechanics [57]. And, in these approaches, the high frequency performance of quantum devices is mainly understood from static quantum simulations. It is assumed that the quantum device behaves as a (small-signal) circuit. The resistances and capacitances of such circuit are then computed from static quantum simulations to evaluate variations of current (conductance) or charge (capacitance) for different voltage.

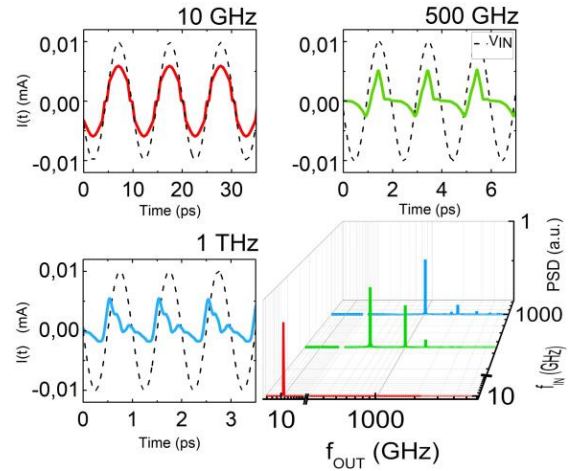

Fig. 6. Total current for a resonant tunneling diode as a function of time for different input frequencies $f$ of a small-signal input voltage (dashed black in arbitrary units). In bottom left, power spectral density (PSD) as a function of the output frequency for the three currents, confirming (nonlinear) harmonic generation [63].

Fortunately, a direct quantum modelling of the displacement current in quantum devices without either the (small-signal) circuit assumption or the perplexing effects of the "collapse" law is possible. There are quantum theories where electrons have well-defined properties independently of their measurement (observation). Such quantum theories without observers, for example Bohmian mechanics [59,60], are well-known in the community dealing with the foundations of quantum mechanics. These remain mostly ignored in the electron device community. Yet, the great advantage of Bohmian formulation of quantum phenomena is that the evaluation of the displacement current can be done following a similar strategy used in the semi-classical Monte Carlo simulations [47,61,62]. Once quantum (Bohmian) trajectories $x(t)$ are computed, satisfying the continuity equation (2), the computation of the displacement follows straightforwardly without any need of the orthodox "collapse". In Fig. 6, we plot the total current (particle plus displacement current) computed from Bohmian trajectories as a function of time for a resonant tunneling diode biased on a sinusoidal signal at different frequencies. The frequency-dependent non-linear behavior of the total current can be related to memory effects. There is plenty of room for unexplored applications of tunneling devices working at frequencies higher than the inverse of the electron transit time, where displacement current becomes more important than the particle current [63].

## V. DISCUSSION

By now, it should be clear to the reader that there are two major threads running through this article. Unfortunately, both tend to be ignored and not sufficiently appreciated within the engineering community. In some sense, the first actually explains the second. The first is the historical record of just when displacement current entered into the world of electricity and magnetism. Here, it seems clear that it was Kirchhoff who first introduced this displacement some 4 years prior to Maxwell. Nevertheless, it was Maxwell who showed the importance in the connection with magnetism as well as currents. The second thread running through this work is the importance of displacement current, in that it forces the consideration of time-varying events into electromagnetics. Without this time varying term, there would be no wave equations for use in fields ranging from electric power distribution to optical information processing. There is also a caveat that comes with this importance of time variation, and this is that phrases such as "dielectric constant" are an oxymoron. The dielectric function is never constant except over very narrow frequency ranges.

Even in what is known as linear response [64], the dielectric function of a simple material like a semiconductor is a very complicated (even nonlinear) function of frequency with multiple poles and zeroes [65], and it is further complicated by the formation of excitons, the existence of band-gap narrowing, and other dynamic effects. The method of studying this dielectric function is spectroscopy, which was discussed above. In composite systems, this becomes much harder to accomplish. It is clear that such a simple system as an ion channel of Figure 5 is an enormously complicated compound system, with each atom or molecular structure having its own dielectric response. Determining the overall dielectric response is extremely difficult and challenges our level of understanding at the fundamental level [45,47].

Even in modern semiconductor devices, layers of thin film materials are stacked and adjoined to one another. Even with simple stacking of thin films, such as in growth of superlattices and heterostructures, determination of the dielectric response, even over a limited range of frequency, is difficult [66]. By the time one tries to couple single photons to single quantum dots embedded into photonic bandgap material, the task is almost impossible [67]. Even the optical dielectric response of a single semiconductor (or even metal) layer is governed by the valence (bonding) electron response to the a.c. signals, and this is usually in the ultra-violet spectral region. Determination of the temporal response in this region, and the delay in which the electrons follow the optical signal, has fallen to the use of attosecond laser pulses [68].

There is a beautiful moral that arises from the discussion of this section. It seems that Kirchhoff was motivated by the continuity equation in (2) to postulate the origin of the displacement current. Despite the fact that the concept of an electron was not known at that time, the meaning of Kirchhoff's law (3), or the continuity equation (2), indicates that the electrons leaving a volume are equal to the those entering minus the temporal variation on the electrons inside such volume. By Gauss' law, such temporal variation of the charge generates displacement current. Kirchhoff's intuition at the middle of the 19$^{th}$ century is the seminal work for the development of our information society. After almost two centuries, his intuition is still helpful in computing displacement current, even in modern quantum electron devices working at THz frequencies, allowing the never-ending progress of our information society. Indeed, the importance of the displacement current addition to Ampere's Law, and the continuity of the total current, provide the important elements necessary to design and understand our circuits for today's technology and provide striking insights into many other systems of daunting complexity.

## REFERENCES

[1] P. Horowitz and W. Hill, *The Art of Electronics*, 3rd Ed., Cambridge: Cambridge Univ. Press, 2015

[2] C.A. Balanis, *Advanced Engineering Electromagnetics*, Hoboken: John Wiley, 2012

[3] S.H. Hall and H.L. Heck, *Advanced signal integrity for high-speed digital designs*. Hoboken: John Wiley, 2011

[4] H.W. Johnson and M. Graham, *High-speed signal propagation: advanced black magic*. New York: Pearson, 2003

[5] E.B. Joffe and K.-S. Lock, *Grounds for Grounding*., Piscataway: (Wiley-)IEEE Press, 2010

[6] V. Bush, *Operational Circuit Analysis*, New York: Chapman & Hall, 1929

[7] E.A. Guillemin, *Communications Networks: Vol. 1 The Classical Theory of Lumped Constant Networks*. New York: John Wiley, 1931

[8] E.A. Guillemin, *Theory of Linear Physical Systems: Theory of physical systems from the viewpoint of classical dynamics, including Fourier methods*, New York: Dover Publications, 2013

[9] O. Heaviside, *Electromagnetic Theory*, London: The Electrician, 1893.

[10] For example, one could go directly to the Wikipedia page for the topic to see this.

[11] P. C. Becker *et al*., "Femtosecond Photon Echoes from Band-to-Band Transitions in GaAs," *Physical Review Letters*, vol. 61, 1647-1649 (1988).

[12] J. Zheng and M. C. Trudeau, *Handbook of Ion Channels*, Boca Raton: CRC Press, 2015.

[13] T.K. Simpson, *Maxwell on the Electromagnetic Field: A Guided Study*, New Brunswick: Rutgers Univ. Press, 1998

[14] J. C. Maxwell, "On Physical Lines of Force. III. The Theory of Molecular Vortices Applied to Statical Electricity," *Philosophical Magazine*, Ser. 4, vol. 23, 11-23, April-May 1861, doi.org/10.1080/14786431003659180

[15] G. Kirchhoff, "Ueber die Bewegung der Elektricität in Leitern," *Annalen der Physik Pogg.*, vol. 102, 193-217 (1857); Tr. as G. Kirchhoff, "On the Motion of Electricity in Wires," *Philosophical Magazine*, Ser. 4, vol. 13, 393-412 (1857).

[16] P. Graneau and A. K. T. Assis, "Kirchhoff on the Motion of Electricity in Wires," *Apeiron*, vol. 19, 19-25 (1994).

[17] H. C. Ørsted, *Experimenta circa effectum conflictus electrici in acum magneticam*. Copenhagen: P. O. Pederson, 1820.

[18] A. M. Ampère, "Du Mémoire sure l'Action mutuelle entre deux courans electriques, entre un courant électrique et un animant o le globe Terrestre, et entre deux aimans," *Annales de Chimie et dePhysique,* vol. 15, 170-218 (1820).

[19] A. M. Ampère, *Exposé des Nouvelles Découvertes sur L'Magnétisme et*

L'Électricité. Paris: C. Méquignon-Marvis, 1822.

[20] A. K. T. Assis and J. P. M. C. Chaib, *Ampère's Electrodynamics*. Montreal: C. Roy Keys, 2015.

[21] A. M. Ampère, *Mémoire sur la theorie mathématique des phénomènes électro-dyanmiques uniquement déduite de l'expérience.* Paris: Bachelier, 1825.

[22] J. C. Maxwell, *A treatise on Electricity and Magnetism*, vol. 2. Oxford: Clarendon Press, 1881, p. 163.

[23] M. Faraday, "Experimental Researches in Electricity," *Philosophical Transactions of the Royal Society*, vol. 122, 125-162 (1832).

[24] M. Faraday, "Experimental Researches in Electricity—Twenty-eighth Series," *Philosophical Transactions of the Royal Society*, vol. 142, 25-56 (1852).

[25] M. Faraday, "On the Physical Character of the Lines of Magnetic Force," Philosophical Magazine, Ser. 4, vol. 3, 401-428 ( June 1952).

[26] W. Weber, "Electrodynamische Maafsbestimmungen," Annalen der Physik, vol. 73, 193-240 (1848); Tr. in "On the Measurement of Electro-dynamic Forces," *Scientific Memoirs*, Ed. by R. Taylor, London: Taylor and Francis, 1852, pp.489-552

[27] J. C. Maxwell, "On Faraday's Lines of Force," *Transactions of the Cambridge Philosophical Society*, vol. 10, 155-229 (1855).

[28] W. Thomson, "Theory of Magnetic Induction in Crystalline and Non-Crystalline Substances," *Philosophical Magazine*, Ser. 4, vol. 1, 177-186, 1851.

[29] J. C. Poggendorf, "Uber die Leitung galvanischer Ströme durch Flussigkeiten," *Annalen der Physik*, vol. 64, 54-57 (1845).

[30] F. E. Neumann, "Allgemeine Gesetze der inducirten elektrischen Ströme," *Annalen der Physik*, vol. 70, 31-44 (1846).

[31] G. Kirchhoff, "Naftrag zu dem Afsatze: Uber den Durchgang eines Elecktrischen Ströme durch ein Ebene, inbesondere durch eine kreisformige," *Annalen der Physik*, vol. 64, 497-514 (1845).

[32] G. Kirchhoff, "Uber ein Problem bei Lineare Verzweigung ecktrischen Ströme," *Annalen der Physik*, vol. 67, 273-283 (1845).

[33] G. Kirchhoff, "Uber the Auflösung der Gleichungen, auf welchen man bei der Untersuchung der linearen Verteilungen galvanischer Ströme geführt wird," Annalen der Physik, vol. 72, 497-508 (1847); Tr. by J. B. O'Toole, "On the Solutions of the Equations Obtained from the Investigation of the Linear Distribution of Galvanic Currents," *IRE Transactions on Circuit Theory*, vol. 9, 4-7 (March 1958).

[34] G. Kirchoff, "Uber die Anwendbarkeit der Formeln für die Intensitäten der galvanischen Ströme in eine Systeme linearer Leiter auf Systeme, die zum Theil aus nicht linearen Leiter bestehen," *Annalen der Physik*, vol. 75, 189-206 (1848).

[35] G. Kirchoff, "Uber eine Ableitung der Ohm'schen Gesetze, welches sich an die Theorie die Elektrostatik anschliefst," Annalen der Physik, vol. 75, 506-513 (1849)

[36] G. Kirchoff, "Bestimmung der Constanten von welcher die Intensität induciter elektrisher Ströme abhängt," Annalen der Physik, vol. 76, 412-426 (1849)

[37] A. Liénard, *Champ électrique et magnétique produit par une charge électrique concentrée en un point et animée d'un movement quelconque*. Paris: G. Caree et C. Naud, 1898, pp. 16, 53, 106.

[38] E. Wiechert, "Elektrodynamische Elementare gesetz," *Annalen der Physik*, Ser. 4, vol. 4, 667-689 (1901).

[39] G. Kirchhoff, "Uber die Bewegung der Elektricität in Leitern," *Annalen der Physik*, vol. 102, 529-542 (1857).

[40] J. C. Maxwell, "A Dynamical Theory of the Electromagnetic Field," *Philosophical Transactions of the Royal Society*, vol. 155, 459-512 (1865).

[41] R. Eisenberg, "A Necessary Addition to Kirchhoff's Current Law of Circuits," *archiv.org*, https://doi.org/10.31224/2234.

[42] B. O. Seraphim and N. Bottka, "Field Effect of the Reflectance in Silicon," Physical Review Letters, vol. 15, 104-107 (1965).

[43] A. N. Gorlyak, I. A. Khramtsovsky, and V. M. Solonukha, "Ellipsometric Application in Optics of Inhomogeneous Media," *Scientific and Technical Journal of Information Technologies, Mechanics and Optics*, vol. 15, 378-386 (2015).

[44] E. Barsoukov and J. R. Macdonald, *Impedance Spectroscopy: Theory, Experiments, and Applications*, New York: John Wiley, 2018.

[45] A. P. Demchenko, Ultraviolet Spectroscopy of Proteins, Berlin: Springer, 2013.

[46] J. Leuthold, C. Koos, and W. Freude, "Nonlinear Silicon Photonics," *Nature Photonics*, vol. 4, 535-544 (2010).

[47] R. Eisenberg, X. Oriols, and D. K. Ferry, "Dynamics of Current, Charge, and Mass," *Molecular Based Mathematics in Biology*, vol. 5, 78-115 (2017).

[48] J. F. Scott, "Ferroelectrics go Bananas," *Journal of Physics: Condensed Matter*, vol. 20, 021001 (2008).

[49] M. Ismail *et al.*, "Photovoltaic Response of 'Ferroelectric' Bananas," *Europhysics Letters*, vol. 125, 47001 (2019).

[50] M. Kline and I. W. Kay, *Electromagnetic Theory and Geometrical Optics*, New York: John Wiley, 1965.

[51] V. Radisic *et al.*, "Power Amplification at 0.65 THz Using InP HEMTs," *IEEE Transactions on Microwave Theory and Techniques*, vol. 60, 724-729 (2012).

[52] A. L. Hodgkin and R. D. Keynes, "The Potassium Permeability of a Giant Nerve Fiber," *Journal of Physiology*, vol. 128, 61-88 (1955).

[53] B. Hille, *Ion Channels of Excitable Membranes*, Oxford: Oxford University Press (2001).

[54] A. L. Hodgkin and A. F. Huxley, "A quantitative description of membrane current and its application to conduction and excitation in nerves," *Journal of Physiology*, vol. 117, 500-544 (1952).

[55] A. Mirenenko et al., "The Persistent Question of Potassium Channel Permeation Mechanisms," *Journal of Molecular Biology*, vol. 433, 167002 (2021).

[56] R. Landauer, "Conductance from Transmission: Common Sense Viewpoints," *Physica Scripta*, vol. T42, 110-114 (1992).

[57] X. Oriols and D. K. Ferry, "Why Engineers are Right to Avoid the Quantum Reality Offered by the Orthodox Theory?" Proceedings of the IEEE, vol. 109, 955-961 (2021).

[58] X.Oriols and D.K.Ferry "Quantum transport beyond DC," Journal of Computational Electronics, vol. 12(3), 317-330 (2013).

[59] D. Bohm, "A Suggested Interpretation of the Quantum Theory in Terms of 'Hidden' Variables," *Physical Review*, vol. 85, 166-179 (1952).

[60] Xavier Oriols and Jordi Mompart, *Applied Bohmian Mechanics: from naoscale systems to cosmology,* 2nd edition, Singapore: Jenny Stanford Publishing , 2019.

[61] X. Oriols, "Quantum-trajectory approach to time-dependent transport in mesoscopic systems with electron-electron interactions" Physical Review Letters, vol. 98, 066803 (2007).

[62] D. Marian, N. Zanghi, and X. Oriols "Weak Values from Displacement Currents in Multiterminal Electron Devices " Physical Review Letters, vol. 116, 110404 (2016).

[63] M. Villani *et al.*, "There is Plenty of Room for THz Tunneling Electron Devices Beyond the Transit Time Limit" IEEE Electron Device Letters, vol. 42(2), 224-227 (2021).

[64] J. Lindhard, "On the Properties of a Gas of Charged Particles," *Danske Matematisk-Fysiske Meddelelser*, vol. 28, 1-57 (1954)

[65] D. K. Ferry, *Semiconductors* (New York: Macmillan, 1991) Ch. 12

[66] G. R. L. Sohie and G. N. Maracas, "In-situ Measurement and Control of Growth Parameters in Molecular Beam Epitaxy," *Proceedings IEEE Conference on Control Applications*, vol. 1, 539-543 (1994).

[67] P. Lodahl, S. Mamoodian, and S. Stobbe, "Interfacing Single Photons and Single Quantum Dots with Photonic Nanostructures," *Review of Modern Physics*, vol. 87, 347-400 (2015)

[68] M. Th. Hassan *et al.*, "Optical Attosecond Pulses and Tracking the Nonlinear Response of Bound Electrons," *Nature*, vol. 530, 66-70 (2016).